\documentclass[12pt]{iopart}
\usepackage{epsfig}

\usepackage{amsfonts, amssymb}
\newcommand{\be}{\begin{equation}}
\newcommand{\ee}{\end{equation}}
\newcommand{\bea}{\begin{eqnarray}}
\newcommand{\eea}{\end{eqnarray}}

\def\fun#1#2{\lower3.6pt\vbox{\baselineskip0pt\lineskip.9pt
\ialign{$\mathsurround=0pt#1\hfill##\hfil$\crcr#2\crcr\sim\crcr}}}

\newcommand\eq[1]{Eq.~(\ref{#1})}

\newcommand\gsim{\mathrel{\rlap{\lower4pt\hbox{\hskip1pt$\sim$}}
    \raise1pt\hbox{$>$}}}

\def\dslash{\not{\hbox{\kern-2pt $\partial$}}}
\def\Dslash{\not{\hbox{\kern-4pt $D$}}}
\def\Oslash{\not{\hbox{\kern-4pt $O$}}}
\def\Qslash{\not{\hbox{\kern-4pt $Q$}}}
\def\pslash{\not{\hbox{\kern-2.3pt $p$}}}
\def\kslash{\not{\hbox{\kern-2.3pt $k$}}}
\def\qslash{\not{\hbox{\kern-2.3pt $q$}}}

 \newtoks\slashfraction
 \slashfraction={.13}
 \def\slash#1{\setbox0\hbox{$ #1 $}
 \setbox0\hbox to \the\slashfraction\wd0{\hss \box0}/\box0 }

\def\ee{\end{equation}}
\def\be{\begin{equation}}

\begin{document}
\setlength{\unitlength}{1mm}

\title[A Constraint on Planck-scale Modifications to Electrodynamics with CMB
polarization data]{A Constraint on Planck-scale Modifications to Electrodynamics with CMB polarization data}

\author{Giulia Gubitosi, Luca Pagano, Giovanni Amelino-Camelia, Alessandro Melchiorri
\footnote[1]{giulia.gubitosi@roma1.infn.it\\
luca.pagano@roma1.infn.it\\
giovanni.amelino-camelia@roma1.infn.it\\
alessandro.melchiorri@roma1.infn.it}}
\address{Physics Department, University of Rome ``La
Sapienza'', \\ and Sezione Roma1 INFN\\P.le Aldo Moro 2, 00185 Rome, Italy}
\author{Asantha Cooray\footnote[2]{acooray@uci.edu}}
\address{Center for Cosmology, Dept. of Physics \& Astronomy, University of
California Irvine, Irvine, CA 92697}

\begin{abstract}
We show that the  Cosmic Microwave Background (CMB) polarization data gathered by the BOOMERanG 2003
flight and WMAP
provide an opportunity to investigate {\it in-vacuo} birefringence, of a type expected in
some quantum pictures of space-time, with a sensitivity that extends even beyond
the desired Planck-scale energy.
In order to render this constraint more transparent we rely on a well studied phenomenological
model of quantum-gravity-induced birefringence, in which one easily establishes
that effects introduced at the Planck scale would amount to values of a dimensionless
parameter, denoted by $\xi$, with respect to the Planck energy
which are roughly of order $1$. By combining
BOOMERanG and WMAP data we estimate  $\xi \simeq -0.110 \pm 0.076$ at the $68\%$ c.l.
Moreover, we  forecast on the sensitivity to $\xi$ achievable
by future CMB polarization experiments (PLANCK, Spider, EPIC), which, in the absence of
systematics, will be at the $1$-$\sigma$ confidence of $8.5 \times 10^{-4}$
 (PLANCK), $6.1 \times 10^{-3}$ (Spider), and $1.0 \times 10^{-5}$  (EPIC) respectively.
The cosmic variance-limited sensitivity from CMB is $6.1\times 10^{-6}$.
\end{abstract}


\maketitle


\section{Introduction}
The challenge of finding a theory for ``quantum gravity" to reconcile quantum mechanics with
the General Relativity description of gravitational phenomena
has been confronting the theoretical physics community for more than 70 years~\cite{stachelhisto}.
A reason why after many decades we understand little on
the ``quantum-gravity problem" originates from the difficulties encountered in experimentally accessing
the realm of quantum corrections to gravity, expected  to be originating
 at the  ultrahigh energy scale given by the Planck energy, $E_p  \sim 10^{28}$eV (or equivalently
the ultrashort Planck length, $L_p \equiv 1/E_p \sim 10^{-35}$m).

 Over the last decade the search for experimental manifestations of Planck-scale effects
 has been reenergized by the realization that certain types of observations in astrophysics
 do provide indirect access to certain Planck-scale effects~\cite{grbgac,gampul,billeretal,kifune,gacNATURE2000eGACPIRANprd,jaconature}.
 Several authors have already argued that
 cosmological observations may soon also help in the efforts of searching for hints on the realm of
 quantum gravity. These expectations originate from the fact that many cosmological observations
 reflect the properties of the Universe at very early times, when the typical
 energies of particles were significantly closer to the Planck scale than the energies presently
 reached in our most advanced particle accelerators. Moreover,
 the particles studied in cosmology have typically travelled ultra-long (``cosmological") distances,
 and therefore even when they are particles of relatively low energies they could be affected
 by a large accumulation of the effects of the ``space-time quantization", which is one of the
 most common expectations emerging from quantum-gravity research\footnote{Most approaches
 to the quantum-gravity problem predict some
form of space-time quantization, although different approaches often lead to profoundly different
predictions on what space-time quantization should entail.}.

Over the last few years some authors (see, {\it e.g.}, Refs.~\cite{qgpCosm1,qgpCosm2,qgpCosm3,qgpCosm4})
have indeed started to use ideas coming from quantum-gravity research
in cosmology, but the possible connections between cosmological measurements and quantum-gravity
research have yet to be fully appreciated in the literature. This is because in the relevant cosmological studies, while one indeed easily sees that the nature of a
certain effect is of interest for constraining models of quantum gravity, it is often not easy to establish whether the conjectured
magnitude of the change could plausibly be connected with corrections introduced genuinely
at the Planck scale. Our main objective here is to establish that in the case of studies
of quantum-gravity-induced {\it in-vacuo} birefringence, proposed in several quantum-gravity
studies (see, {\it e.g.}, Refs.~\cite{gampul,Myers2003,gacNewJournPhys}) with the
CMB polarization data we may  indeed reach sufficient sensitivity to test effects introduced
 at the Planck energy scale.

Several studies have already investigated
(see, {\it e.g.}, Refs.\cite{Xia:2008si,chinapol,cabella,durrer2008,Komatsu:2008hk,Finelli:2008jv})
the relevance of CMB polarization data for testing birefringence, and some studies
have motivated the analysis
also from a broadly-intended quantum-gravity perspective~\cite{durrer2008,Kostelecky2007},but there is still
no dedicated study attempting to establish whether the sensitivity levels provided by CMB observations
could provide access to birefringence introduced at the Planck energy scale.
Here, we make this connection clear and
provide a constraint using existing BOOMERanG and WMAP
polarization data. Moreover, we also make predictions for the
expectations with future CMB polarization measurements both with space and sub-orbital experiments.

In order to make our case for Planck-scale sensitivity more transparent we model birefringence
in the context of a much-studied formalization of Planck-scale modifications of electrodynamics,
a model proposed by  Myers and Pospelov~\cite{Myers2003}, with Lagrangian density of the form
\begin{equation}
 \mathcal L=-\frac{1}{4}F_{\mu\nu}F^{\mu\nu}+\frac{\xi}{2E_{p}} \varepsilon^{jk l}
  F_{0 j} \partial_0
 F_{kl} \, .
\label{myepos}
\end{equation}
This model was proposed as a field theory formalization of the intuition
on Planck-scale modifications of the energy-momentum dispersion relation
that had been developed in a series of preceding studies starting from those
in Refs.~\cite{grbgac,gampul,mexweave}.
The Lagrangian has been much studied from the quantum-gravity
perspective, including several analyses on its observable features
(see, {\it e.g.} Refs.~\cite{jaconature,gacNewJournPhys,Gleiser2001,mattinglyLRR,gacLRR,liberati0805}).

We feel that the fact that our suggestion for testing Planck energy scale corrections in
CMB polarization data is based on such a well-studied model
should render our findings more significant. Moreover, in the case of
the Myers-Pospelov field theory it is relatively simple to establish
which range of values of the single parameter of the model, $\xi$, would amount to introducing
the effects at the Planck energy scale: the correction term in (\ref{myepos})
is of dimension 5, so that the parameter $\xi$ must be dimensionless,
and several studies~\cite{jaconature,Myers2003,Gleiser2001,mattinglyLRR,gacLRR,liberati0805}
have shown that quantization of space time at the Planck scale
would lead to correction terms  with $\xi$ roughly of order 1 ($|\xi| \sim 1$)
in equation~(\ref{myepos}).
This estimate of course must be handled prudently, since the arguments estimating that
the scale of quantum-gravity effects should be given by the Planck scale allow a certain
range of uncertainty: for example, most of these arguments rely crucially
on the rather optimistic assumption of having no running of the gravitational
coupling constant and no role for gravitational effects in the running of other
coupling constants, but plausible estimates of how the renormalization-group flow might
be affected by gravitational effects show~\cite{gacLRR,wilczekGUPeEplanck} that the
quantum-gravity scale may well differ by 2 or 3 orders of magnitude from the Planck scale.
It is therefore rather significant that
the analysis we report here establishes robustly that available
CMB polarization data are sensitive to $|\xi| \sim 1$ and that planned more refined measurements
of the CMB polarization will provide access to values of $|\xi|$ even a few orders
of magnitude smaller than 1.

Interestingly, through a combined analysis of BOOMERanG and WMAP data as we describe in detail later
we find the following estimate: $\xi \simeq -0.110 \pm 0.076$ at the $68 \%$ c.l..
This constraint very explicitly shows that indeed the relevant CMB polarization data have
sensitivity at Planck-scale energies. And it is also intriguing
that the analysis leads us to an estimate of $\xi$
which is nonzero at roughly the $1.5 \sigma$ level.
We shall stress that most previous analyses of birefringence in relation
 to BOOMERanG and WMAP data assume that the polarization angle rotation, usually denoted
 by $\alpha$, is energy independent, $\alpha = \alpha_0$,
 whereas within the Myers-Pospelov framework one finds a characteristic energy
 dependence of $\alpha$: $\alpha = \frac{2 \xi}{E_{p} } E^2 T$, where $E$ is
 the energy of the radiation and $T$ is the time of propagation\footnote{This dependence is inverse to the energy dependence expected
in polarization rotation associated with Faraday rotation
in the presence of a magnetic field with the rotational angle proportional to $E^{-2}$.}.
Using the same data that lead to our result $\xi =-0.110 \pm 0.076$ one would instead
 find $\alpha_0 = (-2.4 \pm 1.9)^o$ assuming an energy independent rotation.
 One could therefore (however weakly) argue that the  evidence in favor of
 birefringence is somewhat more robust
 for the case of the energy-dependent effect we have considered (a $1.5\sigma$ effect versus
 a $1.3\sigma$ effect).

We also comment on previous attempts to constrain $\xi$ using astrophysical data,
such as the ones reported in Refs.~\cite{jaconature,Gleiser2001,liberati0805}.
These analyses provide bounds on $\xi$ that naively appear to be tighter than the
ones we obtained here. However, we find that the frame dependence of the analysis
of such symmetry-breaking effects renders those astrophysical analyses inapplicable
to the context of CMB studies, performed of course in the reference frame naturally
identified by the CMB. A crucial point for this observation is the realization
that the space-isotropic form  of the Myers-Pospelov Lagrangian density can at best
be assumed in a very restricted class of frames of reference: even when the Lagrangian density
takes space-isotropic form in one class of frames it will not be in general space-isotropic
in other frames.
And our reasoning leads us also to conclude that in the
 hypothetical scenario of future more sensitive CMB polarization data
 providing more robust evidence of energy-dependent birefringence (say at a $3 \sigma$ level),
it would be inappropriate to rush to the conclusion that the
space-isotropic Myers-Pospelov framework is finding confirmation. In such a scenario
one should then necessarily perform searches of possible evidence of anisotropy.

The paper is organized as follows: In the next Section, we describe
the Myers-Pospelov framework,
and its implications for birefringence.
Then in Section~3 we discuss some aspects of the phenomenology
of the Myers-Pospelov framework that are particularly relevant
for studies of the CMB.
Section~4
contains our key result obtained from CMB polarization observations.
In Section~5 we compare our analysis to previous, partly related, analyses, and in particular
we discuss the issue of frame-dependent spatial isotropy.
We conclude, in section~6, with a brief summary of our results and some observations
on the outlook of the studies we are here proposing.
We shall work throughout adopting ``natural units"
($\hbar=c=1$), and conventions $\eta^{\mu\nu}=\mathtt{diag}(1,-1,-1,-1)$, for the
metric, and $\varepsilon_{123}=1$, $\varepsilon^{123}=-1$, for the Levi-Civita symbol.

\section{Effective field theory for Planck-scale-modified electrodynamics}
Motivated by previous studies~\cite{grbgac,gampul,mexweave} which had discussed
some mechanisms for Planck-scale  modifications
of the laws of propagation of microscopic particles,
in Ref.~\cite{Myers2003} Myers and Pospelov proposed to describe
such effects within the framework of low-energy effective field theory,
and observed that,
assuming essentially only that the effects are linear in the (inverse of the) Planck scale
and are characterized mainly by an external four-vector, one arrives
at a single possible correction term for the Lagrangian density of electrodynamics:
\begin{equation}
 \mathcal L=-\frac{1}{4}F_{\mu\nu}F^{\mu\nu}
 +\frac{1}{2E_p} n^\alpha F_{\alpha\delta}n^\sigma
  \partial_\sigma(n_\beta\varepsilon^{\beta\delta\gamma\lambda}F_{\gamma\lambda})
 \label{eq:D5lagrangian}
\end{equation}
where the four-vector $n_\alpha$ parameterizes the effect.

The (dimensionless)
components of $n_\alpha$ of course take different values in different reference frames,
transforming indeed as the components of a four-vector.
The quantum-gravity intuition that motivates the study of this model also inspires the
choice of the factor $\frac{1}{E_p}$, and leads one to the expectation that
the components of the Myers-Pospelov four-vector should (when nonzero) take values
somewhere in the
range $10^{-3}  < n_\alpha < 10^{3}$. Values of order $1$ would essentially correspond to
introducing the effects exactly at the Planck scale. Values as high as $10^3$ would still be plausible,
especially in light of some renormalization-group arguments
suggesting~\cite{wilczekGUPeEplanck} that the characteristic scale of quantum-gravity effects might
actually be closer to the particle-physics ``grand-unification scale" than to the Planck scale.
There is no robust argument that would suggest that the characteristic scale of
quantum-gravity effects  might instead
be much higher than the Planck scale, but most authors
prudently consider values of parameters such as the $n_\alpha$'s
still meaningful even down to $10^{-3}$, because of the desire
to be rather cautious before excluding the possibility
of a quantum-gravity interpretation.

These arguments provide motivation for developing a phenomenology
of the Myers-Pospelov field theory exploring a four-dimensional parameter, $n_\alpha$,
keeping in focus the most interesting range of values, $10^{-3}  < n_\alpha < 10^{3}$,
and contemplating the characteristic frame dependence of the parameters $n_\alpha$.
There is already a rather sizeable literature on this phenomenology (see, {\it e.g.},
Refs.~\cite{mattinglyLRR,gacLRR,liberati0805} and references therein)
but still fully focused on what turns out to be the simplest possibility
for the Myers-Pospelov framework, which is the one of assuming to be
in a reference frame where $n_\alpha$
only has a time component, $n_\alpha = (n_0,0,0,0)$.
Then, upon introducing the convenient notation  $\xi \equiv (n_0)^3$,
one can  rewrite (\ref{eq:D5lagrangian}) as
\begin{equation}
 \mathcal L=-\frac{1}{4}F_{\mu\nu}F^{\mu\nu}+\frac{\xi}{2E_{p} }
  \varepsilon^{jkl} F_{0 j} \partial_0F_{k l}\, ,
 \label{eq:MP}
\end{equation}
and in particular one can exploit the simplifications provided
by spatial isotropy.

We shall also focus here on this case $n_\alpha = (n_0,0,0,0)$, although, unlike previous authors,
we shall be rather careful (particularly in Subsection~5.1)
in assessing the limitations that this simplifying assumption
introduces.

From (\ref{eq:MP}) one easily obtains
modified Maxwell equations for the electric and magnetic fields:
\begin{eqnarray}
 0&=&\partial_j E^j\nonumber\\
0&=&-\partial_0 E^k -\partial_j \varepsilon^{ljk}B_l-\frac{2 \xi}{E_{p} }  (\partial_0)^2  B^k, \label{eq:EOMBE}
\end{eqnarray}
and
\begin{eqnarray}
0&=&  \partial_k \varepsilon^{0k j l } E_l-\partial_0 B^j\nonumber\\
0&=&\partial_j B^j. \label{eq:EOMhomogeneous}
\end{eqnarray}
 Note that this second set of equations is
undeformed with respect to the classical ($\xi=0$) case, since it follows only from the antisymmetry of the electromagnetic tensor $F_{\mu\nu}$.
The resulting equation of motion for the electric field is:
\begin{equation}
0=-\partial^0\partial_0 E^k -\partial^l (\partial_l E^k-\partial^k  E_l)-\frac{2\xi}{E_{p} }   \varepsilon^{m k n }( \partial_0)^2  \partial_m  E_n.\label{eq:EOME}
\end{equation}
For the case of  plane waves, $E_m(x)=E_m(k)e^{-ik^\rho x_\rho}$, this equation of motion takes the form:
\begin{equation}
0=k_0^2 E^j + k^l (k_l E^j-k^j  E_l)-\frac{2 i \xi}{E_{p} }   ( k_0)^2  (  k_m \varepsilon^{m j n } E_n) \, .\label{eq:EOMEk}
\end{equation}

Since the Lagrangian density of \eq{eq:MP} is still symmetric under space-rotations there is
clearly no loss of generality in fixing the direction of propagation of the plane wave in, say,
the direction $\hat z$, with wave vector  $k_\rho=(\omega,0,0,p)$.
For such a plane wave propagating along the $\hat z$ axis the three
equations of motion given by \eq{eq:EOMEk} take the form
\begin{eqnarray}
0&=&\omega^2 E_3\nonumber\\
 0&=&-\omega^2 E_1 +p^2 E_1+\frac{2 i \xi}{E_{p} }   \omega^2   p\,  E_2 \nonumber\\
0&=&-\omega^2 E_2 +p^2 E_2-\frac{2 i \xi}{E_{p} }   \omega^2   p\, E_1.
\end{eqnarray}
The first of these equations simply states that the longitudinal component of the electric field is still absent (even for $\xi\neq 0$), and therefore for a plane wave propagating along the $\hat z$ axis one has $E_3=0$. The remaining two equations can be compactly reorganized as follows:
\begin{equation}
 0 =\left(-\omega^2  +p^2 \right)\left(E_1\pm i E_2\right)\pm\frac{2  \xi}{E_{p} } \omega^2   p\, \left(E_1\pm i   E_2  \right),
\end{equation}
meaning that the right- and left- circularly polarized fields $\vec E_\pm\equiv E^1 \hat\epsilon_x \pm i E^2\hat \epsilon_y$ satisfy the equation of motion:
\begin{equation}
\left(-\omega^2  +p^2 \pm\frac{2  \xi}{E_{p} }   \omega^2 p   \right) \vec E_\pm =0\label{eq:E_pmdisprel}.
\end{equation}
So the electric field is subject to birefringence: the left- and right-circular polarized components have different dispersion relations:
\begin{equation}
 \omega_\pm\approx p\left(1 \pm  \frac{\xi}{E_{p} } p \right),
\end{equation}
where we included only the leading correction in $\xi/E_{p} $. The components
propagate with different group velocities $v_{g\pm}= \frac{\partial E_\pm}{\partial p}\approx 1 \pm 2\frac{\xi}{E_{p} } p$.

Assuming that the electric field is originally linearly polarized, then, due to the
different group velocities of its circularly polarized
components, if the time of propagation of the radiation is sufficiently long,
one might have an observably large difference in the time of arrival of
the two circularly-polarized components of the field, and the linear polarization
would therefore be lost \cite{Gleiser2001}.
If instead the time of propagation is
not sufficient to produce a  detectable
difference in the time of arrival of the two modes
with different group velocities, the direction of
the original linear polarization is rotated (see \cite{Kostelecky2002} and references therein).
In particular, if the wave propagates for a time $T$, the amount of rotation is:
\begin{equation}
 \alpha = (\omega_--\omega_+)T = 2 \frac{\xi}{E_{p} } p^2 T \, . \label{eq:deltaalpha}
\end{equation}

\section{Effects on CMB polarization and its power spectrum}
CMB radiation is known to have been generated with some degree of linear polarization (see e.g. \cite{Kosowsky:1994} and references therein), and some linear polarization is observed, so our analysis must be performed in the regime governed by \eq{eq:deltaalpha}.
In light of \eq{eq:deltaalpha},  measurements of the amount of rotation of the CMB  polarization  can be interpreted as measurements of the parameter $\xi$.
To be more precise, one has to consider the fact that the CMB photons propagate into an expanding universe, so the energy of the photons is redshifted during the propagation. The photon energy redshift dependence is:
\begin{equation}
 \omega(z)=\omega_0(1+z),
\end{equation}
where $\omega_0$ is the energy of the photon when it arrives on Earth.
So the amount of rotation of the polarization plane is, for a CMB photon propagating for a time $t$ from the last scattering surface ($t=0$):
\begin{equation}
 \alpha(t)=\int_0^t  2 \frac{\xi}{E_{p} } p(t')^2  dt',\label{eq:deltaalpharedshifttime}
\end{equation}
The amount of rotation  of the polarization of photons propagating from the last scattering surface toward us   ($t=T$) can be written in terms  of the redshift $z$ as (on the last scattering surface $z=Z$, now $z=0$):
\begin{equation}
\alpha=2 \frac{\xi}{E_{p} }p_0^2 \int_0^Z   (1+z) H^{-1}  dz.
\end{equation}
Since to a first approximation $H = H_0 \sqrt{\Omega_m(1+z)^3+\Omega_\Lambda}$, where $H_0\simeq 10^{-18}s^{-1}$ is the Hubble parameter (in terms of the reduced Hubble constant $H_0= h\cdot 100 (Km/s)/Mpc $):
\begin{equation}
\alpha=2 \frac{\xi}{E_{p} }\frac{p_0^2}{H_0} \int_0^Z   \frac{(1+z)}{\sqrt{\Omega_m(1+z)^3+\Omega_\Lambda}}   dz.\label{eq:deltaalpharedshift}
\end{equation}

The formal description of the rotation with angular velocity $\omega\equiv \frac{d\alpha}{d \eta}= \frac{d\alpha}{d t}\frac{a}{a_0}$ of the CMB polarization plane  due to any physical effect requires a modification  of the Boltzmann equation which governs the time evolution of a single Fourier mode of the perturbation of the Stokes parameters $Q$ and $U$. For scalar perturbations the Boltzmann equations become \cite{Liu2006, Kosowsky1996}:
\begin{eqnarray}
 \dot\Delta_Q+i k \mu  \Delta_Q&=&-\dot \tau\left[\Delta_Q+\frac{1}{2}(1-P_2(\mu))S_P\right]+2\omega \Delta_U\nonumber\\
\dot\Delta_U+i k \mu  \Delta_U&=&-\dot\tau\Delta_U-2\omega\Delta_Q \label{eq:boltzmann}
\end{eqnarray}
where the dots indicate derivatives with respect to the conformal time $\eta$, $\mu$ is the cosine of the angle between the photon direction and the Fourier wave vector, $\dot\tau\equiv x_e n_e \sigma_T a/a_0$, with $x_e$ the ionization fraction, $n_e$ the electron number density, $\sigma_T$  the Thomson cross-section and $a$ the scale factor ($a_0$ the scale factor today). $S_p\equiv \Delta_{T2}+\Delta_{Q2}-\Delta_{Q0}$ is the source function,  $\Delta_{Ti}$ and $\Delta_{Qi}$ indicate, respectively, the $i$-th moment in the expansion of the temperature perturbation $\Delta_T$ and of the polarization perturbation $\Delta_Q$ in Legendre polynomials; $P_2(\mu)$ is the second Legendre polynomial.
Note that  the equation for $\Delta_U$ has no generating term, since the $U$ component of polarization is not generated by scalar perturbations,  but by tensors.

These coupled differential equations can be separated  taking the sum and the difference of the first equation with $i$ times the second.
Then a formal integration leads to the solution \cite{Liu2006}:
\begin{equation}
 (\Delta_Q\pm i\Delta_U)(\eta_0)=\int_0^{\eta_0}d\eta\left[ \dot\tau(\eta) e^{i k\mu (\eta-\eta_0)-\tau(\eta)}  S_p(\eta) e^{\pm 2 i \int_{\eta_0}^{\eta} d\eta' \omega(\eta')} \right] \label{eq:exactdelta}
\end{equation}
where $\tau(\eta)=  \int^{\eta_0}_{\eta} d\eta' \dot \tau(\eta')$ and $\eta_0$ is the conformal time today.

Since in our case both $\omega$ and $\alpha$ are expected to be very small and the visibility function $\dot\tau(\eta)e^{-\tau(\eta)}$  peaks tightly at the time of decoupling $\eta=\eta_{dec}$, a very good approximation to the above solution is given by:
\begin{eqnarray}
 \Delta_Q(\eta_0)&=&\tilde \Delta_Q(\eta_0)  \cos\left(2 \int_{\eta_0}^{\eta_{dec}} d\eta' \omega(\eta')\right) = \tilde \Delta_Q(\eta_0)  \cos\left(2 \alpha \right)  \nonumber \\
  \Delta_U(\eta_0)&=&\tilde \Delta_Q(\eta_0) \sin\left(2 \int_{\eta_0}^{\eta_{dec}} d\eta' \omega(\eta')\right)=-\tilde \Delta_Q(\eta_0) \sin\left(2 \alpha\right)
\end{eqnarray}
where $\tilde \Delta_Q(\eta_0)\equiv \Delta_Q(\eta_0)|_{\omega\equiv 0}$ and $\alpha$ is given by eq. (\ref{eq:deltaalpharedshift}).
Taking into account the tensorial modes of the perturbations, which generate a primordial $U$ mode, it can be shown that the effects of a rotation of the polarization
direction are:
\begin{eqnarray}
 \Delta_Q(\eta_0)&=&  \tilde \Delta_Q(\eta_0)  \cos\left(2 \alpha \right)+ \tilde \Delta_U(\eta_0)  \sin\left(2 \alpha \right)   \nonumber \\
  \Delta_U(\eta_0)&=&-\tilde \Delta_Q(\eta_0) \sin\left(2 \alpha\right)+\tilde \Delta_U(\eta_0) \cos\left(2 \alpha\right) \, .
\end{eqnarray}

This leads to the polarization power spectrum $C_\ell^{XY} \sim \int dk\;\left[ k^2 \Delta_X(\eta_0)\Delta_Y(\eta_0)\right] $, $X,Y = T,E,B$:
\begin{eqnarray}
 C_\ell^{EE}&=&\tilde C_\ell^{EE}  \cos^2\left(2 \alpha\right) +\tilde C_\ell^{BB}  \sin^2\left(2 \alpha\right)  \nonumber \\
C_\ell^{BB}&=&\tilde C_\ell^{EE}  \sin^2\left(2 \alpha\right)  +\tilde C_\ell^{BB}  \cos^2\left(2 \alpha\right) \nonumber \\
C_\ell^{EB}&=&\frac{1}{2}\left(\tilde C_\ell^{EE}-\tilde C_\ell^{BB}\right)  \sin\left(4 \alpha\right)     \nonumber \\
C_\ell^{TE}&=&\tilde C_\ell^{TE}  \cos\left(2 \alpha\right)     \nonumber \\
C_\ell^{TB}&=&\tilde C_\ell^{TE}  \sin\left(2 \alpha\right)     \label{eq:powerspectra}
\end{eqnarray}
 We checked numerically that evaluating the power spectra given by equation~(\ref{eq:powerspectra}) instead of  the exact formula in equation~(\ref{eq:exactdelta}) (and the corresponding one for tensorial modes), leads to an error on each $C_\ell^{XY}$ that, for $\int_{\eta_0}^{\eta_{dec}} d\eta' \omega(\eta')$ up to six degrees, is much less than $1\%$.

\section{Data Analysis}

In this section we discuss the current constraints on $\alpha$ and on $\xi$ from the most recent CMB polarization datasets coming from the WMAP five-year mission \cite{Komatsu:2008hk,Dunkley:2008ie} and the BOOMERanG polarization flight in 2003 \cite{2005astro.ph..7503M}.
We follow the  standard approach in the literature by making use of the the publicly available Markov Chain Monte Carlo
package \texttt{cosmomc} \cite{Lewis:2002ah} with a convergence
diagnostics done through the Gelman and Rubin statistic.

In our analysis we consider the complete set of power spectra (\ref{eq:powerspectra}), taking into account also $TB$ and $EB$ cross-correlation power spectra, which are usually set to zero in the standard CMB calculations. We sample the following eight-dimensional set of cosmological
parameters, adopting flat priors on them: the physical baryon and Cold Dark Matter
densities, $\omega_b=\Omega_bh^2$ and $\omega_c=\Omega_ch^2$, the ratio of the sound
horizon to the angular diameter distance at decoupling, $\theta_s$, the scalar spectral index
$n_s$, the overall normalization of the spectrum $A_s$ at $k=0.05$
Mpc$^{-1}$, the optical depth to reionization, $\tau$, the tensor to scalar
ratio of the primordial spectrum, $r$ and, finally, the rotation of
the power spectrum discussed above, $\alpha$. We derive the value of $\xi$ for each Markov chain step using (\ref{eq:deltaalpharedshift}), so including the dependence on the geometrical parameters of the universe. Simultaneously
we also use a top-hat prior on the age of the Universe as $10 {\rm Gyr} < t_0 < 20$ Gyr.
Furthermore, we consider purely adiabatic initial conditions, we impose
flatness, we treat the dark energy component as a cosmological constant, and we include the weak lensing effect in the CMB spectra computation.

{\bf
}
In Table \ref{tab:results1} we show the current constraints on $\alpha$ and $\xi$ from current experiments.
WMAP 5-year data provide the constraint of $\alpha= (-1.6\pm2.1)^o$ (consistent with  previous analyses  \cite{Komatsu:2008hk, Xia:2008si}),
i.e. with no indication for $\alpha$ different from zero, while BOOMERanG data give $\alpha= (-5.2 \pm 4.0)^o$, with a $\sim 1.3 \sigma$ hint for $\alpha <0$.
Assuming $\alpha$ energy-independent ($\alpha=\alpha_0$), the joint estimation from BOOMERanG and WMAP5 data gives  $\alpha_0= ( -2.4 \pm 1.9)^o$, which is also consistent with \cite{Xia:2008si}.
As regards to the $\xi$ parameter, from WMAP 5-year data we constrain it to be $\xi=-0.09 \pm 0.12$, while with BOOMERanG data\footnote{For the analysis using the BOOMERanG
dataset, we fixed the optical depth $\tau$ to the fiducial value of $\tau = 0.09$.} we obtain $\xi=-0.123 \pm 0.096$.
Since $\xi$ is energy-independent, we can also make a combined analysis of the two datasets, which results in the estimation of $\xi=-0.110\pm0.075$.

\begin{table}[!htb]
\begin{center}
\begin{tabular}{rc|c}
Experiment &$\alpha \pm \sigma(\alpha)$ &$\xi \pm \sigma(\xi)$\\
\hline
WMAP ($94$ GHz)& -1.6 $\pm$ 2.1 & -0.09 $\pm$ 0.12 \\
BOOMERanG ($145$ GHz)& -5.2 $\pm$ 4.0 & -0.123 $\pm$ 0.096\\
WMAP+BOOMERanG& - & -0.110 $\pm$ 0.075
\end{tabular}
\caption{Mean values and $1\sigma$ error on $\alpha$(in degrees) and $\xi$ for WMAP, BOOMERanG and WMAP+BOOMERanG}
\label{tab:results1}
\end{center}
\end{table}

We also analyze the ability of future experiments in constraining $\alpha$ and $\xi$. We created several mock datasets with noise properties consistent
 with the PLANCK satellite \cite{:2006uk}, the Spider balloon borne experiment \cite{Crill:2008rd} and the EPIC satellite \cite{Bock:2008ww}, assuming, as fiducial model, the best fit of WMAP with $\alpha=0$ (see Table.\ref{tab:exp}
for the experimental specifications used in the analysis).

We analyze these datasets with the full sky exact likelihood routine \cite{Lewis:2005tp} in the cosmomc code
 including the $f_{sky}^2$ factor to reduce the degrees of freedom,
 ignoring correlations between different multipoles $\ell$.
Firstly, we consider the specifications for the PLANCK satellite
 for the channels at $70$, $100$, $143$ and $217$ GHz
respectively. Clearly the presence of different channels in the PLANCK experiment
has the main goal of foreground removal. However we report the sensitivity on
$\alpha$ and $\xi$ for each channel considering  the possibility of a frequency
dependence of the cosmic signal. As we can see in Table \ref{tab:results2}
the PLANCK channel with the highest sensitivity on $\alpha$ is the $143$ GHz channel,
able to constrain $\alpha$ at the level of $0.07 ^o$ and $\xi$ at level of $0.0017$.
Lower sensitivities on $\alpha$ will be reached by other ``high frequency'' channels but,
 because of the energy dependence of $\alpha$, the $217$ GHz channel is the most sensitive to $\xi$, since it allows to constrain this parameter at the level of $0.0010$. The LFI $70$ GHz channel can provide a constraint of the order
of $0.64^o$ for $\alpha$ and $0.06$ for $\xi$. It is interesting to observe that combining the three HFI channels ($100,\;143,\;217$ GHz) it is possible to improve the sensitivity on the $\xi$ parameter, up to $8.5 \cdot 10^{-4}$.

Before continuing it is important to bear in mind that any systematic present in the calibration of the polarization angle will be reflected
in a spurious $\alpha$ and, as a consequence, a spurious $\xi$.  The current HFI and LFI channels are expected
to be calibrated at a level of about $\sim 1^o$ thanks to  measurements
of polarized emission from the Crab nebula. The expected values reported
in the Table are therefore not taking in to account this possible
systematic. However, a multi-frequency analysis could permit to distinguish a systematic effect due to miscalibration from a true birefringence effect, since the first is energy independent, while  the second has a specific energy dependence.

\begin{table}[!htb]
\begin{center}
\begin{tabular}{rccc}
Experiment & Channel & FWHM & $\Delta T/T$ \\
\hline
PLANCK & 70 & 14' & 4.7 \\
$f_{sky}=0.85$& 100 & 10' & 2.5 \\
& 143 & 7.1'& 2.2\\
& 217 & 5' & 4.8 \\
\hline
Spider & 145 & 40' & 0.30\\
$f_{sky}=0.50$& & & \\
\hline
EPIC& 70 & 12' & 0.066 \\
& 100 & 8.4' & 0.066 \\
& 150 & 5.6' & 0.084 \\
& 220 & 3.8' & 0.17 \\
$f_{sky}=0.85$ & & &
\end{tabular}
\caption{PLANCK, Spider  and EPIC experimental specifications.  Channel
frequency is given
in GHz, FWHM in arcminutes and noise per pixel in $10^{-6}$ for the
Stokes I parameter; the corresponding sensitivities for the Stokes Q and
U parameters are related to this by a factor of $\sqrt{2}$.}
\label{tab:exp}
\end{center}
\end{table}

\begin{table}[!htb]
\begin{center}
\begin{tabular}{rccc}
Experiment & Channel & $\sigma(\alpha)$& $\sigma(\xi)$ \\
\hline
PLANCK & 70 & 0.64 & $6.0\cdot 10^{-2}$ \\
& 100 & 0.14 &$6.5\cdot 10^{-3}$ \\
& 143 & 0.073 & $1.7 \cdot 10^{-3}$\\
& 217 & 0.10 & $1.0\cdot 10^{-3}$\\
& 100+143+217 & - &$8.5 \cdot 10^{-4}$\\
\hline
Spider & 145 & 0.27 & $6.1\cdot 10^{-3}$ \\
\hline
EPIC & 70 & $2.1\cdot 10^{-3}$& $1.9\cdot 10^{-4}$ \\
 & 100 &$1.8\cdot 10^{-3}$& $7.8\cdot 10^{-5}$ \\
 & 150 &$1.5\cdot 10^{-3}$ & $2.9\cdot 10^{-5}$ \\
 & 220 &$1.2\cdot 10^{-3}$& $1.1\cdot 10^{-5}$ \\
 & 70+100+150+220 & - & $1.0\cdot 10^{-5}$ \\
\hline
CVL & 150 &$6.1\cdot 10^{-4}$ & $1.3\cdot 10^{-5}$\\
 & 217 &$6.1\cdot 10^{-4}$ & $6.1\cdot 10^{-6}$
\end{tabular}
\caption{Expected $1\sigma$ error for PLANCK $70,100,143,217$ GHz,
Spider $145$ GHz, EPIC $70,100,150,220$ GHz and two ideal CVL experiment
at $150$ GHz and $217$ GHz on $\alpha$ (in degrees) and $\xi$.}
\label{tab:results2}
\end{center}
\end{table}

It is interesting to extend the forecast to other future experiments as
Spider and EPIC. Spider, a balloon borne experiment that will fly over
Antarctica in 2012, will constrain variation of $\alpha $ at level
of $ 0.27 ^o$ and of $\xi$ at level of $6.1\cdot 10^{-3}$,
 with a sensitivity competitive with PLANCK.  The
EPIC satellite proposal, or an equivalent next generation CMBpol satellite mission,
can detect deviations as small as $0.0012^o$ for $\alpha$ and
 $1.0\cdot 10^{-5}$ for $\xi$, providing a dramatic
improvement respect to PLANCK and Spider. Again, the channel at higher
frequency provides the best constrain.
Finally we consider an ideal experiment, cosmic variance limited in
anisotropy and polarization measurements. This experiment provides
a sort of fundamental limit to the precision that can be achieved.
We found that in this, ideal, experiment, angles as small as
$\alpha =0.0006^o$ could be, in principle, measured, and we observe in Table \ref{tab:results2}
that this would provide sensitivity to $\xi=1.3\cdot10^{-5}$, for an experiment at $150$,
or even $\xi=6.1\cdot10^{-6}$, for an experiment at $217$ GHz.

\section{Comparison with other results}

\subsection{Comparison with limits previously obtained using observations in  astrophysics}
While ours is the first explicit work aimed at comparing the predictions
of the Myers-Pospelov framework to CMB data,
there have been a few other attempts, such as the ones
reported in Refs.~\cite{Gleiser2001,jaconature,mattinglyLRR,liberati0805},
to place limits on the parameter $\xi$ using certain observations in astrophysics.
The outcome of some of these analyses was simply described in terms
of limits on the parameter $\xi$, as if they were absolute limits,
but we must here stress that, in light of the very explicit frame dependence
of the Myers-Pospelov model,
one  of course cannot establish absolute limits on parameters.
From the analysis of experimental data collected in a certain frame one
can definitely obtain even very robust limits on the values of some of the
components of the four-vector $n_\alpha$ (see Eq.~(\ref{eq:D5lagrangian}))
in that frame, but it is highly nontrivial to then qualify such results
from a frame-independent perspective.

All these previous studies have, like the one we are here reporting,
focused on the implications of a time component for the  Myers-Pospelov
symmetry-breaking four-vector $n_\alpha$, a restriction which
is introduced very explicitly by assuming $n_\alpha = (n_0,0,0,0)$.
But of course the quantum-gravity arguments~\cite{grbgac,gampul,mexweave}
 that motivated the study
of the  Myers-Pospelov framework do not provide any indication
that the four-vector should take this form in any specific frame ($n_\alpha$
may well not even be time-like),
and even if it took this form in a certain class of frames it would then definitely
have different form in other frames, transforming like a four-vector from frame to frame.

Constraints on all four components of $n_\alpha$
in one frame could be converted, through appropriate Lorentz transformations,
into analogous (but possibly very different in magnitude)
constraints applicable in another frame. But bounds established exclusively
for the time component of  $n_\alpha$ are of very limited applicability
in other frames.

This is particularly significant for our results since
some analyses of data in astrophysics can lead to very stringent
bounds on $\xi$.
In one recent such study~\cite{liberati0805} it was even argued that observations of polarized radiation
from the Crab Nebula
can be used to obtain the impressive bound $|\xi| <  10^{-9}$, which would
amount to the constraint $|n_0| <  10^{-3}$ on the time component of the four-vector $n_\alpha$.
Clearly these results obtained in astrophysics start to provide us an intuition
that large values of  $n_0$
 are disfavored,
but at the same time it should be noticed that it is not easy to convert
such results into intelligible constraints applicable in reference frames that
are different from the ``laboratory frame" where the bound was derived.
In principle one could even contemplate the possibility that
in such a laboratory frame one has, say, $n_\alpha = (0,10^3,10^3,10^3)$,
which would be compatible with any upper bound on $n_0$ and would still be
meaningful from a quantum-gravity perspective (since, as mentioned in Section~2,
components of $n_\alpha$ with values of order $10^3$ could be expected in
rather popular scenarios estimating the scale of quantum-gravity as roughly
given by the grand-unification scale).
And if indeed $n_\alpha = (0,10^3,10^3,10^3)$ in some laboratory frame then
a boost of unimpressive magnitude, with, say, $\beta \simeq 10^{-3}$,
could take us from that laboratory frame to a frame where the time component
of  $n_\alpha$
is actually of order $1$.

In light of the rather significant frame dependence implied by these considerations
it is rather clear that for this research program it is a top priority
to move on to bounding experimentally  all
four components of $n_\alpha$. And for works that focus on the time component
it appears that it is rather advantageous to focus on studies of the CMB since one can at least
rather easily
combine different sets of data in a meaningful way, by describing them all in
the natural reference frame of the CMB.

Also significant from the perspective of the frame dependence of the Myers-Pospelov
framework is the fact that
the results of our analysis, while being inspired by this framework,
depend most strongly on the assumption of a characteristic (quadratic) energy dependence of $\alpha$,
and not very sensitively on the details of the Myers-Pospelov framework with $n_\alpha = (n_0,0,0,0)$.
Such an energy dependence is a generic feature of any quantum-gravity-inspired
description of birefringence, since it is obtained using only
 dimensional analysis and the fact
that the effects must disappear in the $E_p \rightarrow \infty$ limit.
One should therefore expect a similar energy dependence also
in the Myers-Pospelov framework with generic $n_\alpha$
and in models where the breakdown of Lorentz symmetry is governed by, say, a two-index tensor.

\subsection{Comparison with previous CMB limits on energy-dependent birefringence}
While ours is the first study to discuss a robust link from the parameters of a model
that has been extensively studied in the quantum-gravity literature and CMB polarization
data, from a broader phenomenological perspective
the possibility of a quadratic energy dependence of $\alpha$, which is indeed the main characteristic of
the effects we studied, had already been considered in Ref.~\cite{Kostelecky2007}.
In order to compare of our results to the ones of Ref.~\cite{Kostelecky2007} we must
establish a link between the parameter $\xi$ of the (isotropic) Myers-Pospelov model
and the multi-parameter description of birefringence effects for CMB polarization data analyses
adopted in Ref.~\cite{Kostelecky2007}.
This can be done by describing the
 rotation  of linear polarization of CMB radiation codified in our \eq{eq:deltaalpharedshift}
 in terms of   rotations
of the Stokes vector $\vec s\equiv \left(Q,U,V\right)^T$,
since it is at the level of these rotations of the Stokes vector that
Ref.~\cite{Kostelecky2007} introduced its multi-parameter phenomenological picture.

Let us start by noticing that
the variation of the Stokes vector due to birefringence (not necessarily between circularly polarized components) can be written as \cite{Kostelecky2007}:
\begin{equation}
 \frac{d\vec s}{dt}=2 \omega \vec \zeta\times \vec s \label{eq:dsdt}
\end{equation}
where $\vec \zeta$ is proportional to the Stokes vector of the faster of the two modes subject to birefringence.

 The Stokes vector $\vec \zeta$ can be  written in the spin-weighted
 basis, $\vec \zeta_{SW}\equiv\left(\zeta_{(+2)}, \zeta_{(0)},\zeta_{(-2)}\right)^T\equiv
  \left( \zeta^1-i \zeta^2,\zeta^3,\zeta^1+i\zeta^2\right)^T$,
and each component of $\vec \zeta_{SW}$ can be expanded in spin-weighted spherical harmonics:
\begin{equation}
 \zeta_{(\pm2)}=\sum_{lm} \left(k_{(E)lm}\pm i k_{(B)lm}\right)\,_{\pm 2}Y_{lm},\quad \zeta_{(0)}=\sum_{lm}k_{(V)lm} \,_0Y_{lm}.\label{eq:sphericalexpansion}
\end{equation}

Each coefficient $k_{(V)lm}$  and, respectively, $k_{(E,B)lm}$ in \eq{eq:sphericalexpansion} can be expressed as a combination of only odd and, respectively, even powers of $\omega$.
The parameters $k_{(V)lm}^{(d)}$ and $k_{(E,B)lm}^{(d)}$ used in \cite{Kostelecky2007} are the coefficients of the  expansion: $k_{(V)lm}=\sum_{d\;odd} \omega^{d-4} k_{(V)lm}^{(d)}$, $k_{(E,B)lm}=\sum_{d\;even} \omega^{d-4} k_{(E,B)lm}^{(d)}$.

In the case of our interest, in which the rotation of the Stokes parameter is due to a
birefringence between circularly polarized components of the radiation,
taking conventionally as the fastest mode  the right-circularly polarized one,
one finds that the vector $\vec \zeta$ is proportional to $\left( 0,1,0\right)^T$ in the spin-weighted basis. The variation of the Stokes vector is, in the $(Q,U,V)^T$ basis:
\begin{equation}
 \frac{d\vec s}{dt}=\left(\begin{array}{c} -Q_0 \sin(\alpha)+U_0 \cos(\alpha)\\-Q_0\cos(\alpha)-U_0\sin(\alpha)\\0\end{array}
\right)\frac{d\left(\alpha\right)}{dt},
\end{equation}
where $\alpha(t)$ is given by \eq{eq:deltaalpharedshifttime}: $$\frac{d\alpha}{dt}(t)=2 \frac{\xi}{E_{p} } p(t)^2.$$ Writing this equation in the spin-weighted basis, evaluating it at time $t=T$ (today) and comparing with \eq{eq:dsdt}, one finds  $\zeta_{(0)}=-\frac{\xi}{E_{p} }p_0$.
So all the coefficients of the expansion of $\zeta_{(\pm2)}$ are null, while for $\zeta_{(0)}$
the only nonzero coefficient is $k_{(V)00}=-\frac{\xi}{E_{p} }p_0\sqrt{4\pi}$, because  $\zeta_{(0)}$
doesn't depend on the direction of observation in the sky.
Since $k_{(V)00}$ is linear in the energy of
the photon,
there is only one nonzero coefficient in the expansion of $k_{(V)00}$, and this coefficient,
which was denoted by $k_{(V)00}^{(5)}$ in Ref.~\cite{Kostelecky2007}, is therefore
the only parameter in the parameterization introduced in Ref.~\cite{Kostelecky2007}
that would reflect the presence of a (time-component-only) Myers-Pospelov term in the
Lagrangian density of electrodynamics. The  comparison between our results
and the ones of Ref.~\cite{Kostelecky2007} should therefore be based on
the relationship between our parameter $\xi$ and the parameter $k_{(V)00}^{(5)}$,
which is
\begin{equation}
k_{(V)00}^{(5)}=-\frac{\xi}{E_{p} }\sqrt{4\pi}.
\label{kv005}
\end{equation}
For this parameter Ref.~\cite{Kostelecky2007} arrives at
an estimate $k_{(V)00}^{(5)} \simeq (3\pm2) \times 10^{-20} GeV^{-1}$,
using exclusively BOOMERanG data and without performing a Markov chain analysis,
but rather
fixing the cosmological model to the WMAP best fit model and
neglecting therefore possible correlations between the cosmological
parameters \footnote{An effect of this could be the fact that another
parameter studied by \cite{Kostelecky2007}, $k_{(V)00}^{(3)}$,  is
estimated also from the BOOMERanG dataset to
be  $k_{(V)00}^{(3)}=(12\pm 7)\times 10^{-43} GeV$,
which corresponds to a rotation angle $\alpha=12\pm7$
degrees, while the estimate of $k_{(V)00}^{(5)}=(3\pm2) \times 10^{-20} GeV^{-1}$
from the same dataset corresponds to a rotation angle $\alpha=4\pm3$ degrees.}.
In light of (\ref{kv005}) our result $\xi \simeq -0.110 \pm 0.076$
amounts to $k_{(V)00}^{(5)}=(3.2\pm2.1)\times 10^{-20} GeV^{-1}$.
This is fully consistent with the findings of Ref.~\cite{Kostelecky2007},
and we would like to argue that our result should now be viewed as the benchmark
for the limits on $k_{(V)00}^{(5)}$, both because we relied on a more sizeable data sample,
combining BOOMERanG and WMAP data, and because of the marginalization over the remaining
cosmic parameters.

\section{Conclusions and Outlook}
The main objective of our analysis was to establish robustly that cosmology and
particularly CMB data can provide constraints on effects that are of interest
from a quantum-gravity perspective and are introduced at the Planck scale.
This possibility has been contemplated in several previous studies but never in a way
that would transparently expose sensitivity to effects introduced truly at the Planck scale.
We feel that we have fully achieved our goal
by relying on the Myers-Pospelov model, and on the rather large quantum-gravity literature
on this model, which also provided a clear target for the description of effects
introduced at the Planck scale.
We therefore hope that our analysis will motivate a more intense program of studies
of quantum-gravity effects in cosmology, not necessarily focused on in-vacuo birefringence
and/or the Myers-Pospelov model.

It is intriguing that
our estimate of the Myers-Pospelov $\xi$ parameter, $\xi=-0.110\pm 0.076$, provides a
faint indication for a nonzero effect. In this respect it is certainly valuable
that, as discussed in Section~4, future experiments
will be able to measure CMB radiation polarization
with higher sensitivities, and we find that this should lead
to an improvement of three orders of magnitude
in the constraints obtained following our strategy of analysis ($\sigma(\xi)\sim 10^{-5}$).

As stressed in Subsection 5.1, in light of the rather
significant frame dependence introduced by the Myers-Pospelov setup,
this type of sensitivities obtainable
from CMB-data analyses are still rather meaningful, in spite of the apparently
ultra stringent constraints that have been obtained on the parameter $\xi$
using astrophysical observations.
We also showed that CMB data can be used to determine
the sign of $\xi$, a possibility which was never exposed in previous works
relying on astrophysics and that may prove challenging in those contexts.

From the methodological aspect, while in the past the birefringence of CMB photons
has been  widely studied, and several bounds have already been provided on the
birefringence angle $\alpha$, here we pointed that in the literature the possible
energy dependence of the effect has not been adequately contemplated, leading to
possible misleading conclusions when averaging on different photon frequencies.
We  showed how exploiting the availability of data from different channels of
frequency of the photons improves the sensibility on the birefringence
parameter $\xi$ and leads to a  slightly stronger indication for a
nonzero effect, than the results obtained estimating the birefringence
angle $\alpha$ with an average on frequencies. This should be taken into
account in planning future CMB experiments, and may turn out to be
another reason for combining cosmological and astrophysical investigations
of the Myers-Pospelov framework.

Here, we have analyzed a uniform, isotropic rotation to the EM polarization vectors.
But, as stressed in Subsection 5.1,
the quantum-gravity motivation for investigating the Myers-Pospelov framework and
the intrinsic structure of the Planck-scale correction term added to the Lagrangian density
of electrodynamics provides a strong invitation to contemplate also
effects resulting
in anisotropic rotation of polarization. In an upcoming paper we plan to return
to this issue and address the extent to which CMB polarization measurements
can constrain such an anisotropic model.

\ack
We acknowledge fruitful discussions with Ruth Durrer, Fabio Finelli and Matteo Galaverni.
G.~G., L.~P. and A.~M. are supported by ASI contract I/016/07/0 "COFIS".\\
G.~A.-C. is supported by grant RFP2-08-02 from The Foundational Questions Institute (fqxi.org).

\end{document}